\documentstyle[sprocl]{article}
\input{psfig}
\def\be{\begin{equation}}
\def\ee{\end{equation}}
\def\bea{\begin{eqnarray}}
\def\eea{\end{eqnarray}}

\begin{document}
\title{Disentangling Non-Standard Model $T$-Violating Sources 
in Exclusive Semileptonic $B$ Decays}
\author{ Guo-Hong Wu }
\address{TRIUMF Theory Group, 4004 Wesbrook Mall, Vancouver, B.C. V6T 2A3,
 Canada}
\maketitle\abstracts{
Measurements of the polarization of the lepton and $D^*$ in the 
exclusive semileptonic $B$ decays,
$B \to D \ell \overline{\nu}$ and $B \to D^* \ell \overline{\nu}$,
can be used to separate and identify the non-standard model sources 
of $T$ violation.
These $T$-odd effects, estimated in both supersymmetric and 
non-supersymmetric models, could be within the reach of the planned 
$B$ factories.  
}
\section{Introduction}

The transverse muon polarization in the $K^+_{\mu3}$ decay has long been known
to serve as a probe of non standard model (SM) $T$-violating 
scalar interactions \cite{sakurai,zhit}.
 For the exclusive semileptonic $B$ decays
$B \to D \ell \overline{\nu}$ and $B \to D^* \ell \overline{\nu}$,
  $T$-odd polarization observables (TOPO's) can also be constructed by using
the spin vectors of the lepton and/or vector meson $D^*$.
These observables receive negligible contributions from the 
standard model, and can be used to differentiate between 
the possible structures of non-SM $T$-violating sources \cite{wkn1},
 a feature which is not available for the inclusive semileptonic $B$ 
decays \cite{atwood,grossman}.  
In particular, it will be shown that transverse
polarizations of the $\tau$ lepton in $B \to D$ and $B \to D^*$ decays
are respectively sensitive to phases in the effective scalar and 
pseudoscalar interactions, whereas the $T$-odd $D^*$ polarization
is sensitive to an effective right-handed current effect.
 The Lorentz structures of the new physics can thus be uniquely identified.
 For the purpose of illustration, several classes of models are 
shown to give rise to large $T$-odd effects \cite{wkn2}.

\section{General Analysis}

 We start by deriving the model-independent results for the TOPO's.

\subsection{Four-Fermi Interactions} \label{subsec:ff}

 The physics of semileptonic $B$ decays can be written in 
terms of general four-Fermi interactions,
\begin{eqnarray}
{\cal L} & = & - \frac{G_F}{\sqrt{2}} V_{cb}
        \overline{c} \gamma_{\alpha} (1 - \gamma_5) b
        \overline{\ell} \gamma^{\alpha} (1- \gamma_5) \nu
       +G_S \overline{c} b \overline{\ell} (1 - \gamma_5) \nu
 \nonumber \\
 &&  + G_P \overline{c} \gamma_5 b  \overline{\ell} (1 - \gamma_5) \nu
        + G_R \overline{c} \gamma_{\alpha} (1+\gamma_5)  b
         \overline{\ell} \gamma^{\alpha} (1- \gamma_5) \nu
   + h.c. , \label{eq:interaction}
\end{eqnarray}
where $G_F$ is the Fermi constant, $V_{cb}$ is the relevant
CKM matrix element, and $G_S$, $G_P$ and $G_R$
denote the strengths of the non-SM interactions due to scalar,
pseudoscalar and right-handed current exchange respectively.
Tensor effects are negligible in most cases and will not be considered.

 It is convenient to introduce three dimensionless parameters which
are directly involved in the decay amplitude,
\begin{equation}
\Delta_{S,P}  =   \frac{\sqrt{2}G_{S,P}}{G_FV_{cb}}  
    \frac{m_B^2}{(m_b \mp m_c)m_l}
\;\;\;\;\;\;\;\;
\Delta_R  =  \frac{\sqrt{2}G_R}{G_FV_{cb}},
\end{equation}
where the $-$ and $+$ signs correspond to $\Delta_S$ and $\Delta_P$ 
respectively, and  $m_b$, $m_c$, and $m_l$ are the quark and lepton masses.
The phases in these $\Delta$ parameters could lead to
observable $CP$-violating effects.

\subsection{$\tau$ polarization}
   The transverse lepton polarization can be best studied for the
heaviest lepton, the $\tau$.
It is defined as
\begin{eqnarray}
P^{\bot}_{\tau} & = & \frac{\bf{s}_{\tau} \cdot 
         ( \bf{p}_{D^{(*)}} \times \bf{p}_{\tau})}
{|\bf{p}_{D^{(*)}} \times \bf{p}_{\tau}|},
\end{eqnarray}
where $\bf{s}_{\tau}$ is the spin vector of the $\tau$.
For numerical estimates, we use for the form factors the leading
order results in heavy quark effective theory, and take the Isgur-Wise
function to be $\xi(w)=1.0 - 0.75 \times (w-1)$ as the average value
of current data \cite{xidata}. 

   For the $B\rightarrow D \tau \overline{\nu}$ decay, the transverse 
$\tau$ polarization depends on an effective scalar interaction.
Its value when averaged over the whole phase space is
\begin{eqnarray}
\overline{P^{(D)}_{\tau}} & = & - \overline{\sigma_D} Im \Delta_S =
- 0.22 \times Im \Delta_S \, ,
\label{eq:polavtauD}
\end{eqnarray}
where the coefficient is proportional to the mass ratio $m_{\tau}/m_B$. 

The $\tau$ polarization in the
$B \rightarrow D^* \tau \overline{\nu}$ decay
is sensitive to an effective pseudoscalar interaction.
And its average is
\begin{eqnarray}
\overline{P^{(D^*)}_{\tau}} & = & - \overline{\sigma_{D^*}} Im \Delta_P
= - 0.068 \times Im \Delta_P \, .
\label{eq:polavtauD*}
\end{eqnarray}
Note that $\overline{\sigma_{D^*}}$ is about a factor of three smaller than
$\overline{\sigma_D}$.
This is because effectively only one of the three polarization states
of the $D^*$, the longitudinal polarization, contributes to the
transverse $\tau$ polarization \cite{wkn1,garisto}.

\subsection{$D^*$ polarization}

   We now construct TOPO's for $B \rightarrow D^* \ell \overline{\nu}$ decay
using the $D^*$ polarization vector $\vec{\epsilon}$. Since we are not
measuring the spin of the lepton, we can take the $\ell=e, \mu$ modes
which have a larger branching ratio than the $\tau$ mode.
The TOPO's constructed can be shown \cite{wkn2} to 
have a one-to-one correspondence with 
certain $T$-odd momentum correlations in the four-body final state
$B\to D^*(D\pi, D\gamma) \ell \overline{\nu}$.

Working in the $B$ rest frame, we
denote the three-momenta of the $D^*$ and $\ell$
by ${\bf p}_{D^*}$ and ${\bf p}_{\ell}$, and define three orthogonal
vectors
${\vec{n}}_1  \equiv  \frac{({\bf p}_{D^*} \times {\bf p}_{\ell})
    \times {\bf p}_{D^*}}
     {|({\bf p}_{D^*} \times {\bf p}_{\ell})
    \times {\bf p}_{D^*}|}$,
${\vec{n}}_2  \equiv  \frac{{\bf p}_{D^*} \times {\bf p}_{\ell}}
                     {|{\bf p}_{D^*} \times {\bf p}_{\ell}|}$,
and
${\vec{n}}_3= \frac{{\bf p}_{D^*}}{|{\bf p}_{D^*}|} \frac{m_{D^*}}{E_{D^*}}$.
The vector ${\vec{n}}_3$ has been chosen such that
the constraint $\epsilon^2=-1$ becomes symmetric in the ${\vec{n}}$'s;
i.e.
  $(\vec{\epsilon}\cdot {\vec{n}}_1)^2 +
  (\vec{\epsilon}\cdot {\vec{n}}_2)^2 +
   (\vec{\epsilon}\cdot {\vec{n}}_3)^2=1$.
 It is useful to note that the $D^*$ polarization projection transverse 
to the decay plane,
${\vec \epsilon}\cdot {\vec n}_2$, is $T$-odd, and that
the projections inside the decay plane,
${\vec \epsilon}\cdot {\vec n}_1$ and ${\vec \epsilon}\cdot {\vec n}_3$,
are $T$-even.

  A measure of the $T$-odd correlation involving the $D^*$ polarization
can be defined as
\begin{eqnarray}
P_{D^*} & \equiv &
 \frac{d\Gamma - d\Gamma^{\prime}}{d\Gamma_{total}} =
  \frac{2d\Gamma_{T-odd}}{d\Gamma_{total}} ,
\label{eq:polD*}
\end{eqnarray}
where $d\Gamma^{\prime}$ is obtained by performing a $T$ transformation
on  $d\Gamma$,
 $d\Gamma_{T-odd}$ is the $T$-odd piece in the partial width,
and $d\Gamma_{total}$ is the partial width summed over
 $D^*$ polarizations.

The ${D^*}$ polarization observables can now be shown to depend on 
right-handed current interactions, and to involve two structures
which can be separated.
Simply averaging over the whole phase space keeps only the first
polarization structure,
\begin{eqnarray}
  \overline{P^{(1)}_{D^*}} &\simeq & 0.51 \times
(\vec{\epsilon}\cdot {\vec{n}}_1)(\vec{\epsilon}\cdot \vec{n}_2)
Im \Delta_R .
\label{eq:TOPO1}
\end{eqnarray}
The second polarization structure may be separated out by
making use of a symmetry under the reflection of the lepton energy.
This procedure gives the second TOPO involving the $D^*$ polarization,
\begin{eqnarray}
  \overline{P^{(2)}_{D^*}}
 &\simeq & 0.40 \times
(\vec{\epsilon}\cdot {\vec{n}}_2)(\vec{\epsilon}\cdot \vec{n}_3)
Im \Delta_R .
\label{eq:TOPO2}
\end{eqnarray}
Complementary measurements of both observables may then be used to
provide a consistency check regarding
the possible existence of a right-handed current.

 The main results of this section are summarized in Table~\ref{tab:general}.

\begin{table}[t]
\caption{TOPO's and effective  four-Fermi interactions
\label{tab:general}}
\vspace{0.4cm}
\begin{center}
\begin{tabular}{|c|c|c|c|} \hline  & & &  \\
   & $\rm{Im}\Delta_S$ & $\rm{Im}\Delta_P$ & 
     $~~~~\rm{Im}\Delta_R~~~~$ \\ \hline
$P_{\tau}^{(D)}$  & $\surd$ &  0 & 0   \\
$P_{\tau}^{(D^*)}$ & 0 &  $\surd$  & 0  \\
$P_{D^*}^{(1,2)}$ & 0 & 0 & $\surd$  \\
 \hline
\end{tabular}
\end{center}
\end{table}

\section{Model Estimates}

In Table~\ref{tab:models}, the maximal $T$-odd polarization effects 
are listed in both supersymmetric (SUSY) and non-SUSY models.
The detailed calculations can be found in Ref. \cite{wkn2}.

\begin{itemize}
\item
 It has been noted \cite{wn} that squark generational mixings between
$\tilde{t}_R$ and $\tilde{c}_R$ can give rise to a double enhancement,
due to $m_t$ and due to large mixing, and can lead to large
$T$-violating effects in $B$ decay.

\item
 For $R$-parity breaking models, bounds from flavor changing neutral
current processes and lepton universality
constrain the $T$-odd effects to be at most a few percent.
There is no induced right-handed current (RHC) at tree level.

\item
 In multi-Higgs-doublet models, the bound from $B \to X \tau \overline{\nu}$
does not rule out a large polarization effect. However, there is no
RHC induced at tree level.

\item
  In leptoquark models, large $T$-odd effects are found to be possible
with  effective scalar and pseudoscalar four-Fermi interactions.

\item
  In left-right symmetric models, if one does not impose 
manifest or pseudo-manifest left-right symmetry, the constraint  on the 
$W_L$-$W_R$ mixing will be less stringent. Sizable RHC can then
give rise to a $D^*$ polarization effect as large as $8\%$.

\end{itemize}

\begin{table}[t]
\caption{Contributions to the effective four-Fermi interactions
and to the various TOPO's from SUSY with squark intergenerational mixing,
SUSY with $R$-parity violation, the three Higgs-doublet model (3HDM),
leptoquark models, and left-right symmetric models (LRSM's).
\label{tab:models}}
\vspace{0.4cm}
\begin{center}
\begin{tabular}{|c|c|c|c|c|c|} \hline & & & & &  \\
     & squark mixing  & $R\!\!\!\!/\!\;$ SUSY & 3HDM  &
        Leptoquarks  & LRSM  \\ \hline
$\Delta_S$  & $\surd$  & $\surd$  &  $\surd$  &  $\surd$  & 0  \\
$\Delta_P$  & $\surd$  & $\surd$  &  $\surd$  &  $\surd$  & 0    \\
$\Delta_R$  & $\surd$  & 0  & 0   &  0  & $\surd$   \\ \hline
$|P_{\tau}^{(D)}|$  & 0.35 &  0.05 & $\sim 1$ &  $\sim 1$ & 0 \\
$|P_{\tau}^{(D^*)}|$  & 0.05 & 0.008 &  0.3 &
   0.2 & 0     \\
$|P_{D^*}^{(1)}|$    & 0.02  &  0 & 0 & 0 & 0.08  \\
$|P_{D^*}^{(2)}|$    & 0.016  & 0 &  0 & 0 & 0.06
\\ \hline
\end{tabular}
\end{center}
\end{table}

\section{Summary}
 The TOPO's constructed for the exclusive semileptonic $B$ decays
have been shown to probe separately effective scalar, pseudoscalar,
and right-handed current interactions. They receive sizable 
contributions from a variety class of models, and could be 
detectable at the planned $B$ factories.

\section*{Acknowledgments}
  I would like to thank K. Kiers and J.N. Ng for the fruitful collaborations
on which this talk is based. This work is partially supported by the Natural 
Sciences and Engineering Research Council of Canada.

\end{document}